\documentclass[12pt,aps,prd,showpacs,amsmath,amssymb]{revtex4}
\usepackage{slashed,amsmath,float}
\usepackage{mathrsfs}%花写体小花\mathscr{P}
\usepackage{amsfonts}%花写体大花\mathfrak{P}
\usepackage{graphicx}
\usepackage{color}
\begin{document}
\title{Electromagnetic form factors of $\Lambda_b$ in the Bethe-Salpeter equation approach}

\author{Liang-Liang Liu $^{a}$}
\email{ liu06_04@mail.bnu.edu.cn}
\author{Chao Wang $^{b}$}
\email{chaowang@nwpu.edu.cn}
\author{Ying Liu $^{(c)}$}
\email{yingliubnu@gmail.com}
\author{ Xin-Heng Guo $^{a}$}
\email{ corresponding author. xhguo@bnu.edu.cn}
\affiliation{\footnotesize (a)~College of Nuclear Science and Technology, Beijing Normal University, Beijing 100875, People's Republic of China}
\affiliation {\footnotesize (b)~Center for Ecological and Environmental Sciences, Key Laboratory for Space Bioscience and Biotechnology, Northwestern Polytechnical University, Xi$^\prime$an, 710072, People's Republic of China}
\affiliation {\footnotesize (c)~Beijing No. 20 High school, Beijing 100085, People's Republic of China}

\begin{abstract}
The heavy baryon $\Lambda_b $ is regarded as composed of a heavy quark and a scalar diquark which has good spin and isospin quantum numbers. 
In this picture, we calculate the electromagnetic (EM) form factors of $\Lambda_b$ in the Bethe-Salpeter equation approach.
We find that the shapes of the EM  form factors of $\Lambda_b$ are similar to those of $\Lambda$,  which have a peak at about $\omega=1$, but the amplitudes are much smaller than those of $\Lambda$.
\end{abstract}


\maketitle

\section{Introduction}
The nucleon electromagnetic (EM) form factors describe the spatial distributions of electric charge and current inside the nucleon and they are intimately related to nucleon internal structure. They are not only  important observable parameters but also a vital key to understand the strong interaction \cite{Arrington, Perdrisat}. 
There were a lot of experimental results on EM form factors of baryons \cite{Bourgeois, Jlab, Walk, Arr, Bost, Lung, VoL, Horn, Tade, Cau, Kub, GDK} and mesons \cite{Len, Col, Fra, J.R.Green} during the past two decades.

The EM form factors of $\Lambda$ and $\Sigma$ were calculated in the framework of light-cone sum rule (LCSR) up to twist 6 \cite{YL-L,HMQ} . The authors provided a fit approach to predict the magnetic moment of a hadron. 
The $Q^2$-dependent EM form factors of the $\Lambda$ baryon were obtained, and were fitted by the dipole formula to estimate the magnetic moment of the $\Lambda$ baryon. 
It was found that the magnetic form factor approaches zero faster than the dipole formula with the increase of $Q^2$.
%%%%\textcolor[rgb]{1.00,0.00,0.00}{ This is different from the results in the case of the nucleon from the polarization technic experiments.}

\begin{figure}
  \centering
  \includegraphics[width=9.5cm,angle=-90]{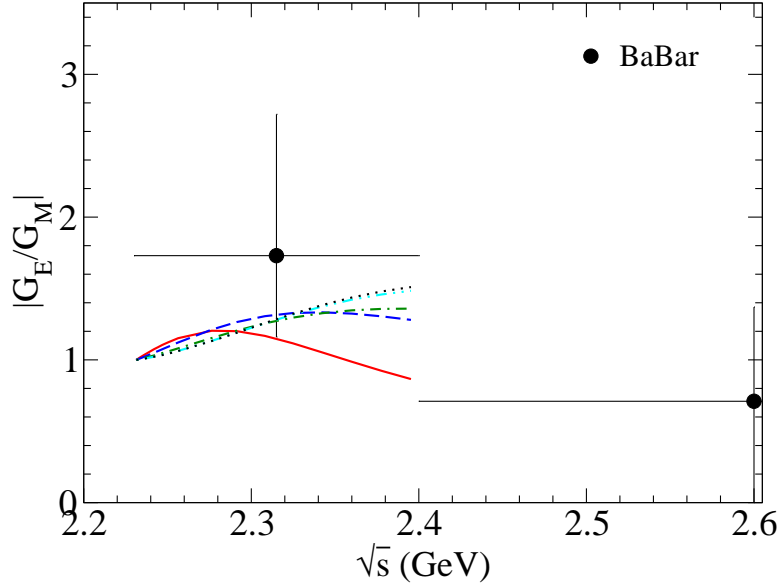}
  \caption{The ratio $|G_E/G_M|$ as a function of the total center-of-mass energy. The solid, dashed, and dash-dotted lines correspond to the $\Lambda \bar{\Lambda}$ models I, II, and II from Ref. \cite{JKV1}, respecticely, the dash-double-dotted and dotted lines correspond to the models K and Q described in Ref. \cite{JKV2} respecticely. (This figure is taken from Ref. \cite{J-U}.)}\label{HMbb}
\end{figure}

\begin{figure}
  \centering
  \includegraphics[width=9.5cm,angle=-90]{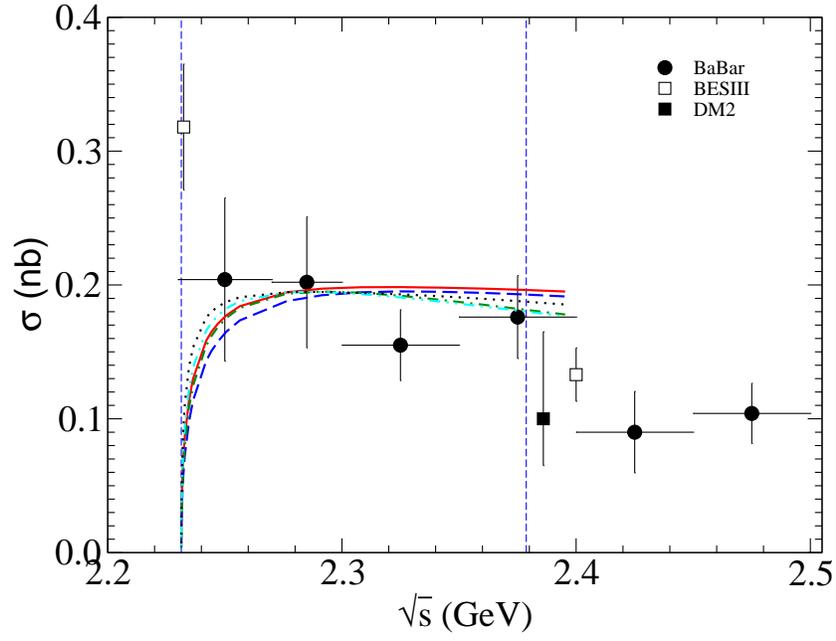}
  \caption{The  total cross section for $ e^+ e^- \rightarrow \Lambda \bar{\Lambda}$. For notation, see Fig. \ref{HMbb}. (This figure is taken from Ref. \cite{J-U}.)}\label{HMl}
\end{figure}

\begin{figure}
  \centering
  \includegraphics[width=9.5cm,angle=-90]{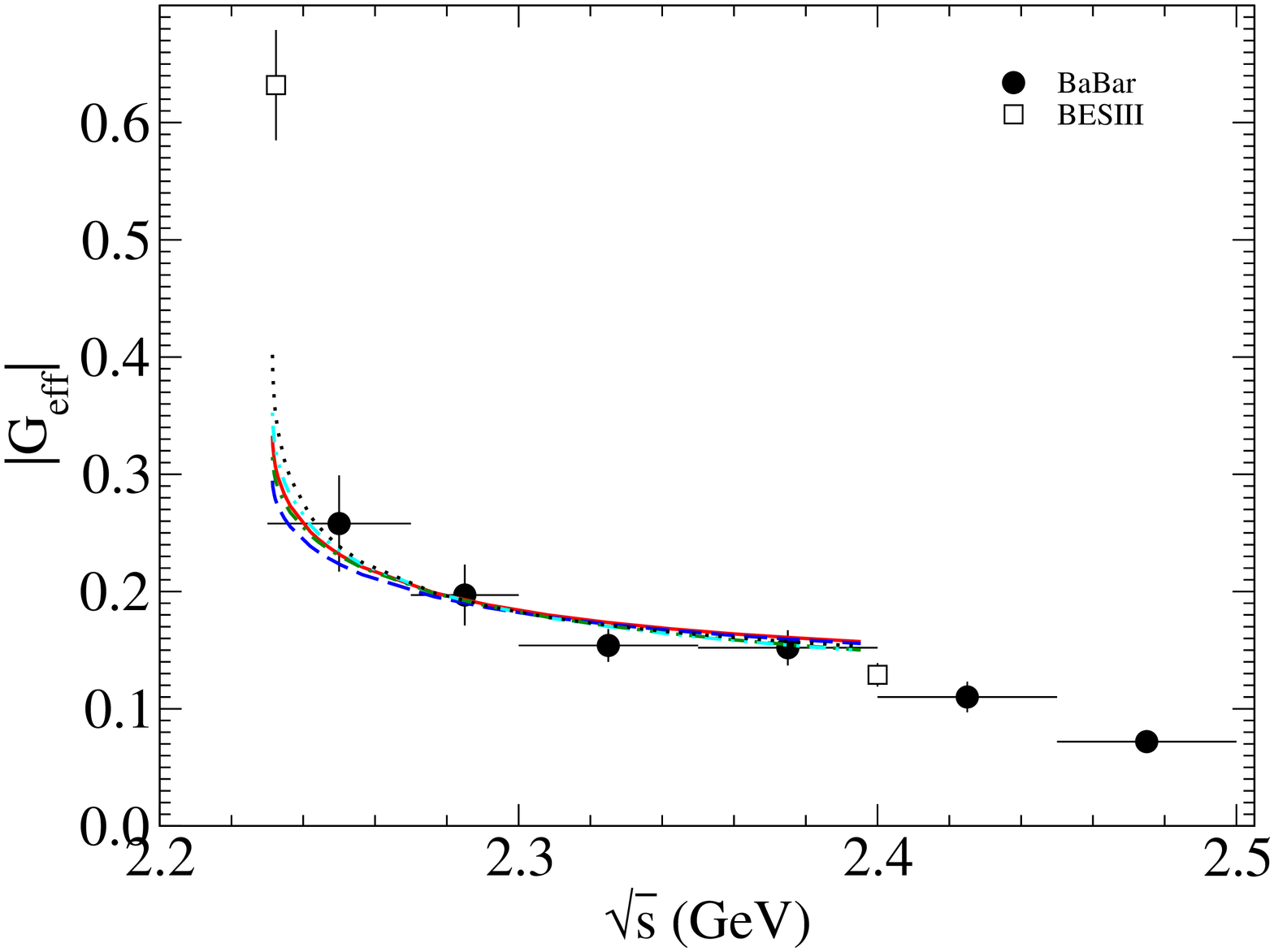}
  \caption{The effective form factor $|G_{eff}|$  for $ e^+ e^- \rightarrow \Lambda \bar{\Lambda}$.  For notation, see Fig. \ref{HMbb}. (This figure is taken from Ref.\cite{J-U}.)}\label{HMr}
\end{figure}

Recently, based on the experiments at BESIII and BaBar \cite{BESIII}, the authors of Ref. \cite{J-U} investigated the process $ e^+ e^- \rightarrow \Lambda \bar{\Lambda}$ in the near-threshold region with specific emphasis on the role played by the interaction in the final $\Lambda \bar{\Lambda}$ state. 
Their calculation was based on the one-photon approximation for the elementary reaction mechanism, and they took into account rigorously the effects of the $\bar{\Lambda} \Lambda$ interaction in close analogy to the work on $e^+ e^−\rightarrow \bar{p}p$ \citep{JXU}.
They gave the form factor ratio $|G_E|/|G_M|$ for $\Lambda$ (Fig. \ref{HMbb}), and found that the form factors ratio is $1$ at the threshold. 
They also gave the  total cross section and effective form factor $|G_{eff}|$ for $ e^+ e^- \rightarrow \Lambda \bar{\Lambda}$ (Figs. \ref{HMl} and \ref{HMr}) with $|G_{eff}|$ being defined as

\begin{eqnarray}\label{geff}
|G_{eff}|=\sqrt{\frac{\sigma_{e^+e^-\rightarrow \bar{\Lambda}\Lambda}}{\frac{4\pi \alpha^2\beta}{3s} \left[1+\frac{2 M^2_{\Lambda}}{s}\right]  }},
\end{eqnarray}
where $s$ is center-of-mass energy, $M_{\Lambda}$ is the mass of $\Lambda$, $\alpha$ is the fine-structure constant and $\beta$ is a phase-space factor. 
The differential cross section can be expressed as the following:

\begin{equation}\label{cross}
\frac{d \sigma}{d \Omega}=\frac{\alpha^2 \beta}{4s}\left[|G_M(s)|^2(1+\cos^2 \theta)+\frac{4 M^2_{\Lambda}}{s} |G_E(s)|^2 \sin^2\theta \right],
\end{equation}
where $\theta$ is the scattering angle in the laboratory frame. As is shown in Figs. \ref{HMl} and \ref{HMr}, the EM form factors of $\Lambda$ have a sudden change when the total center-of-mass energy $\sqrt{s}$ changes in the range $2.2 \sim 2.3$ GeV.

In the present paper we will study the EM form factors of $\Lambda_b$ in the quark-diquark picture.
In this picture, $\Lambda_b$ is regarded as a bound state of two particles: one is a heavy quark and the other is a quasiparticle made of two quarks, or diquark. 
This model has been successful in describing some baryons \cite{H.Meyer,A. De Ruijula,G. Karl,F. Close}.
Since the parity of the $b$-quark is positive, the parity of the diquark involved in the ground state baryon should also be positive. 
Since the isospin of $\Lambda_b$ and the $b$-quark are zero the isospin of the diqaurk (ud) should be zero. Hence the spin of the diquark is also zero. In this picture, the Bethe-Salpeter (BS) equation for $\Lambda_b $ has been studied 
extensively \cite{Zhang-L,Guo-XH, Liu Y, Wu-HK, Weng-Mh}. 
Then $\Lambda_b$ can be described as $b(u d)_{0 0}$ [the first and second subscripts correspond to the spin and the isospin of the $(u d)$ diquark, respectively].
We will calculate the EM form factors in the BS equation approach and compare the results with the EM form factors of $\Lambda$.

The paper is organized as follows. In Section II, we will establish the BS equation for $\Lambda_b $ as a bound state of $b(u d)_{0 0}$.
In Section III we will derive EM form factors for $\Lambda_b$ in the BS equation approach. 
In Section IV the numerical results for the EM form factors of $\Lambda_b $ will be given.
Finally, the summary and discussion will be given in Section V.

\section{BS EQUATION FOR $\Lambda_b$}\label{sec2}

In the previous work \cite{Guo-XH, Zhang-L, Liu Y, Wu-HK}, the BS wave function of $b(ud)_{0 0}$ system is defined as 
\begin{eqnarray}\label{chi-x}
  \chi(x_1,x_2,P) &=& \langle0|T\psi(x_1) \varphi(x_2)|P\rangle,
\end{eqnarray}
where $\psi(x_1)$ is the field operator of the $b$-quark at the position $x_1$, and $\varphi(x_2)$ is the field operator of the scalar diquark at the position $x_2$, $P=M v$ is the momentum of the baryon.
We use $M,~m_q,\text{and}~m_D$ to represent the masses of the baryon, the $b$-quark and the diquark, respectively, and $v$ to represent the baryon's velocity. 
We define the BS wave function in momentum space:
\begin{eqnarray}\label{chi-f}
  \chi(x_1,x_2,P) &=& e^{i P X}\int \frac{d^4 p}{(2\pi)^4}e^{i p x} \chi_P(p),
\end{eqnarray}
where $X= \lambda_1 x_1+\lambda_2 x_2$ is the coordinate of center mass, $\lambda_1=\frac{m_q}{m_q+m_D} $, $\lambda_2=\frac{m_D}{m_q+m_D} $, and $x= x_1-x_2$. As in Refs. \cite{Liu Y,Guo-XH,Wu-HK,Zhang-L}, we can prove that the BS equation for the $b(ud)_{00}$ system has the following form in momentum space:
\begin{eqnarray}\label{chi-p}
\chi_P(p) &=& S_F(p_1)\int \frac{d^4 p}{(2 \pi)^4}K(P,p,q)\chi_P(q)S_D(p_2),
\end{eqnarray}
where $p_1=\lambda_1 P+p$ and $p_2=\lambda_2 P-p$, $K(P,p,q)$ is the kernel that is the sum of all two-particle-irreducible diagrams, $S_F(p_1)$ and $S_D(p_2)$ are propagators of the quark and the scalar diquark, respectively. 
According to the potential model \cite{Guo-XH,E.Eichten}, the kernel is assumed to have the following form:
\begin{eqnarray}\label{kernel}
  -iK(P,p,q) &=&I\otimes I V_1(p,q)+\gamma_\mu \otimes \Gamma^\mu V_2(p,q),
\end{eqnarray}
where $\Gamma^\mu=(p_2+q_2)^\mu\frac{\alpha_{seff}Q_0^2}{Q^2+Q^2_0}$ is introduced to describe the structure of the scalar diquark \cite{Guo-XH,M.Ansel}, and $Q_0^2$ is a parameter that freezes $\Gamma^\mu$ when $Q^2$ is very small. In the high energy region the diquark form factor is proportional to $1/Q^2$, which is consistent with perturvative QCD calculations \cite{GS}. 
By analyzing the EM form factors of the proton, one can take $Q_0^2=3.2GeV^2$ \cite{Zhang-L}.
$V_1$ and  $V_2$ are the scalar confinement and one-gluon-exchange terms that have the following forms in the covariant instantaneous approximation \cite{Guo-XH,Zhang-L,Weng-Mh,Wei-Kw}:
\begin{eqnarray}\label{V1}
  \tilde{V}_1(p_t-q_t) &=& \frac{8 \pi \kappa}{[(p_t-q_t)^2+\varepsilon^2]^2}-(2\pi)^2\delta^3(p_t-q_t)\int \frac{d^3k}{(2\pi)^3} \frac{8 \pi \kappa}{(k^2+\varepsilon^2)^2},
\end{eqnarray}

\begin{eqnarray}\label{V2}
  \tilde{V_2} (p_t-q_t)&=&- \frac{16 \pi }{3}\frac{\alpha_{seff}}{(p_t-q_t)^2+\varepsilon^2},
\end{eqnarray}
where $p_t$ and $q_t$ are the transverse projections of the relative momenta along the momentum $P$ and are defined as $p_t^\mu = p^\mu-p_l v^\mu$ and $ q_t^\mu = q^\mu -q_l v^\mu$ where $p_l=v \cdot p $ and $q_l=v \cdot q$, the second term of $\tilde{V}_1$ is introduced to avoid infrared divergence at the point $ p_t=q_t$, $\varepsilon$ is a small parameter to avoid the divergence in numerical calculations.
The range of the parameter $\kappa$ is $0.02 \sim 0.08~\text{GeV}^3$ \cite{Liu Y,Wu-HK}.

The quark and diquark propagators can be written as the following:
\begin{eqnarray}\label{SF}
  S_F(p_1)&=&\frac{i}{2 \omega_q}\left[ \frac{\slashed{v}\omega_q +(\slashed{p}_t+m_q)}{\lambda_1 M +p_l -\omega_q +i \epsilon}+\frac{\slashed{v}\omega_q -(\slashed{p}_t+m_q)}{\lambda_1 M +p_l +\omega_q -i \epsilon}\right],
\end{eqnarray}  
\begin{eqnarray}\label{SD}
  S_D(p_2) &=&\frac{i}{2 \omega_D} \left[\frac{1}{\lambda_2 M-p_l-\omega_D+i \epsilon}-\frac{1}{\lambda_2M-p_l+ \omega_D-i\epsilon}\right],
\end{eqnarray}
where $\omega_q = \sqrt{m_q^2-p_t^2}~\text{and}~\omega_D = \sqrt{m_D^2-p_t^2} $. Considering $\slashed{v} u (v,s)=u(v,s)$ ($u(v,s)$ is the spinor of $\Lambda_b$ with helicity $s$) , $ \chi_P (p)$ can be written as \cite{Liu Y} 
\begin{eqnarray}\label{chi-pp}
  \chi_P(p) &=& (f_1+f_2 \gamma_5 +f_3 \gamma_5 \slashed{p}_t+f_4 \slashed{p}_t+f_5 \sigma_{\mu \nu}\varepsilon^{\mu\nu\alpha\beta}p_{t \alpha} p_{t \beta}) u(v,s),
\end{eqnarray}
where $f_i ~(i=1,...,5)$ are the Lorentz-scalar functions of $p_t^2$ and $p_l$. Considering the properties of $ \chi_P(p)$ under parity and Lorentz transformations, Equation (\ref{chi-pp}) can be simplified as the following:
\begin{eqnarray}
  \chi_P(p) &=& (f_1+\slashed{p}_t f_2)u(v,s).
\end{eqnarray}

Defining $\tilde{f}_{1(2)}=\int \frac{d p_l}{2 \pi}f_{1(2)}$, we find that the scalar BS wave functions satisfy the coupled integral equation as follows:

\begin{eqnarray}\label{eig1}
\tilde{f}_1(p_t) =&& \frac{1}{4 \omega_D \omega_q(-M + \omega_D+ \omega_q)}\int \frac{d^3q_t}{(2\pi)^3}\{\nonumber\\
&&\left[(\omega_q  +m_q ) (\tilde{V}_1+ 2 \omega_D \tilde{V}_2)-   p _t \cdot ( p _t+ q _t) \tilde{V}_2 \right]\tilde{f}_1(q_t) \nonumber\\
&& + \left[-  (\omega_q+m_q) ( q _t + p _t)\cdot q_t\tilde{V}_2 +  p _t\cdot q_t(\tilde{V}_1- 2 \omega_D \tilde{V}_2)\right]\tilde{f}_2(q_t) \}   \nonumber\\
&&-\frac{1}{4\omega_D \omega_q(M + \omega_D+ \omega_l)} \int \frac{d^3q_t}{(2\pi)^3} \{ \nonumber \\
&&\left[(\omega_q -m_q)(\tilde{V}_1- 2\omega_D \tilde{V}_2)+ 4 p _t\cdot( p _t+ q _t)  \tilde{V}_2 \right]\tilde{f}_1(q_t) \nonumber\\
&& + \left[(m_q- \omega_q )  ( q _t + p _t)\cdot q _t \tilde{V}_2 -   p _t\cdot q _t  (\tilde{V}_1+ 2\omega_D \tilde{V}_2)\right ]\tilde{f}_2(q_t) \},
\end{eqnarray}

\begin{eqnarray}\label{eig2}
\tilde{f}_2(p_t)= && \frac{1}{4\omega_D \omega_q(-M + \omega_D+ \omega_q)} \int \frac{d^3q_t}{(2\pi)^3} \{   \nonumber\\
 &&\left[(\tilde{V}_1+ 2 \omega_D \tilde{V}_2)-( -\omega_q+m_q) \frac{( p _t+ q _t) \cdot p _t }{ p^2_t }\tilde{V}_2\right]\tilde{f}_1(q_t) \nonumber\\
 &&+\left[ \big( (m_q -\omega_q)( \tilde{V}_1+ 2  \omega_D \tilde{V}_2） \big) \frac{ p_t \cdot q_t}{ p^2_t } - (  q^2_t+  p_t \cdot q_t) \tilde{V}_2 \right]\tilde{f}_2(q_t)  \}    \nonumber\\
 &&-\frac{1}{4 \omega_D \omega_q(M + \omega_D+ \omega_q)}\int \frac{d^3q_t}{(2\pi)^3}  \{ \nonumber\\
 && \left[- (\tilde{V}_1- 2\omega_D \tilde{V}_2)+(\omega_q  + m_q)\frac{( p _t+ q _t)\cdot p _t }{ p^2_t } \tilde{V}_2)\right]\tilde{f}_1(q_t)  \nonumber\\
 &&+ \left[\frac{(m_q+\omega_q) (-\tilde{V}_1- 2 \omega_D \tilde{V}_2）) }{  p^2_t }p_t \cdot q_t + (  q^2_t+  p_t \cdot q_t)\tilde{V}_2)\right]\tilde{f}_2(q_t) \}.
\end{eqnarray}

It is noted that the second part of $\tilde{f_i}~(i=1,2)$ in Eqs. (\ref{eig1}, \ref{eig2}) are of order $1/M_{\Lambda_b}$, which is very important to obtain the magnetic form factor. 

In general, the BS wave function can be normalized in the condition of the covariant instantaneous approximation \cite{Liu Y,Wei-Kw}:
\begin{eqnarray}\label{BSNOR}
  i \delta^{i_1 i_2}_{j_1 j_2} \int \frac{d^4 q d^4 p}{(2\pi)^8}\bar{\chi}_P(p,s)\left[\frac{\partial}{\partial P_0}I_p(p,q)^{i_1 i_2 j_2 j_1}\right]\chi_P(q,s^\prime)=\delta_{s s^\prime},
\end{eqnarray}
where $i_{1(2)}$ and $j_{1(2)}$ represent the color indices of the quark and the diquark, respectively, $s^{(\prime)}$ is the spin index of the baryon $\Lambda_b$, $I_p(p,q)^{i_1 i_2 j_2 j_1}$ is the inverse of the four-point propagator written as follows:

\begin{eqnarray}\label{IPNOR}
  I_p(p,q)^{i_1 i_2 j_2 j_1} =\delta^{i_1 j_1}\delta^{i_2 j_2} (2 \pi)^4 \delta^4(p-q)S^{(-1)}_q(p_1)S^{(-1)}_D(p_2).
\end{eqnarray}

\section{EM form factors of $\Lambda_b$}

Generally, the expressions of EM form factors of the spin-$1/2$ baryon B are defined by the matrix element of the EM current between the baryon states \cite{J.R.Green, YL-L,HMQ}:
\begin{equation}\label{JJJ}
\langle B(P',s')|j_\mu(x=0)|B(P,s)\rangle=\bar u(P',s')\left[\gamma_\mu F_1(Q^2)-i\frac{\sigma_{\mu\nu} q^\nu}{2M}F_2(Q^2)\right]u(P,s),
\end{equation}
where $F_1(Q^2)$ and $F_2(Q^2)$ are Dirac and Pauli form factors, respectively, $u(P,s)$ denotes the baryon spinor with momentum $P$ and spin $s$, $M$ is the baryon mass, $Q^2=-q^2=-(P-P')^2$ is the squared momentum transfer, and $j_\mu$ is the EM current relevant to the baryon. 
In particular, for the proton and the neutron the form factors $F_1$ and $F_2$ have the following values  at the point $Q^2 \rightarrow 0$, which corresponds to the exchange of low virtuality photon:
\begin{eqnarray}\label{EMNOR}
  F_{1p(n)}(0) &=& 1(0),   \\
  F_{2p(n)}(0)&=&\kappa_{p(n)},
\end{eqnarray}
where the indices $p$ and $n$ represent the proton and the neutron,  respectively, and $\kappa_p=\mu_p-1$ ($\mu_p$ is the magnetic momentum of the proton), $\kappa_n=\mu_n$ are the anomalous magnetic momenta of the proton and the neutron, respectively. 
In the perturbative QCD theory for the helicity-conserving form factor $F_1(Q^2)$, a dominant scaling behavior at large momentum transfer is predicted \cite{G.P}:

\begin{eqnarray} \label{FH}
  F_1 \sim \left(\frac{1}{Q^2}\right)^{n-1},
\end{eqnarray}
where $n$ is the number of valence quarks in the hadron. The power counting can be justified by QCD factorization theorems which separate short-distance quark-gluon interactions from soft hadron wave functions \cite{GPL,A.Efr,V.L.C,I.G.A,V.A.A,C.E.C}. Hence for a baryon we have 
\begin{eqnarray}\label{F1}
  F_1 \sim  \frac{1}{Q^4}.
\end{eqnarray}

The Pauli form factor $F_2$ requires a helicity flip between the final and initial baryons, which in turn requires, thinking of the quarks as collinear, a helicity flip at the quark level, which is suppressed at high $Q^2$. $F_2$ should have the following bahavior at high $Q^2$ \cite{VCM,ST}:

\begin{equation}\label{F2}
F_2 \sim \frac{1}{Q^6}.
\end{equation}

The Dirac and Pauli form factors are related to the magnetic and electric form factors $G_M(Q^2)$ and $G_E(Q^2)$:
\begin{eqnarray}
G_M(Q^2)&=&F_1(Q^2)+F_2(Q^2), \label{gem1}\\
G_E(Q^2)&=&F_1(Q^2)-\frac{Q^2}{4M^2}F_2(Q^2),\label{gem2}
\end{eqnarray}
where $M$ is the mass of a baryon. At small $Q^2$, $G_E$ and $G_M$ can be thought of as Fourier transforms of the charge and magnetic current densities of the baryon. However, at large momentum transfer this view does not apply. Considering Eqs. (\ref{F1} - \ref{gem2}), at the large momentum transfer $|G_E|/|G_M|$ should be a stable value.

In our present work, we will calculate the EM form factors of $\Lambda_b$. When we consider the quark current contribution we have
\begin{equation}\label{MJJ}
\langle \Lambda_b(v',s')|j_\mu^{quark}|\Lambda_b(v,s)\rangle=\bar u(v',s')[ g_{1q}(Q^2)\gamma_\mu+g_{2q}(Q^2)(v'+v)_\mu ]u(v,s),
\end{equation}
where $j_\mu^{quark}=\bar{b}\gamma_\mu b$, $v^{(')}=P^{(')}/M_{\Lambda_b}$ is the velocity of $\Lambda_b$. 

Define $\omega=v'\cdot v=\frac{Q^2}{2 M^2_{\Lambda_b}}+1$ as the velocity transfer, $g_{1q}$, and $g_{2q}$ become functions of $\omega$ \cite{Guo-XH,Liu Y,G-K}.
 When $\omega=1$, to order $\frac{1}{M_{\Lambda_b}}$, we have the following relation \cite{Guo-XH}:

\begin{eqnarray}\label{Nor}
  g_{1q}(1)+2 g_{2q}(1) &=& 1 +\mathcal{O}(1/M^2_{\Lambda_b}).
\end{eqnarray}

\begin{figure}[!ht]
\begin{center}
\includegraphics[width=9.5cm] {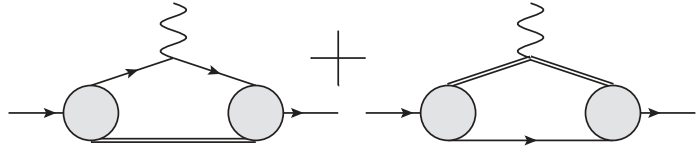}
\caption{The EM current is the sum of the quark current and the diquark current \cite{V.Keiner} }\label{diq}
\end{center}
\end{figure}

In our work, we will use Eq. (\ref{Nor}) to normalize BS wave functions and neglect $1/M_b^2$ corrections \cite{G-K}. This relation has been proven to be a good approximation \cite{G-K} for a heavy baryon and proposed in \cite{M-B,M-V,H-M,BJM} for mesons.

In the quark-diquark model, the electromagnetic current $j_\mu$ coupling to $\Lambda_b$ is simply the sum of the quark and diquark currents, see Fig. \ref{diq}. So we have the relation  \cite{J.R.Green}:
\begin{eqnarray}\label{jj}
  j_\mu &=& j_\mu^{quark}+ j_\mu^{diquark},
\end{eqnarray}
where $j_\mu^{diquark}= \bar{D}\Gamma_\mu D $, $\Gamma_\mu$ is the vertex among the photon and the  diquark which includes the scalar diquark form factor. Hence, we have 
\begin{equation}\label{g1g2}
\langle \Lambda_b(v',s')|j_\mu |\Lambda_b(v,s)\rangle=\bar u(v',s')[ g_{1 }(Q^2)\gamma_\mu+g_{2 }(Q^2)(v'+v)_\mu ]u(v,s).
\end{equation}

Comparing Equations (\ref{g1g2}) and (\ref{JJJ}), we have:
\begin{eqnarray} 
g_{1}&=&F_1-\frac{F_2}{2},\label{g1q} \\
g_{2}&=&\frac{F_2}{4}.\label{g2q}
\end{eqnarray}

It can be shown that the matrix elements of the quark current and the diquark current can be written as the following:
\begin{eqnarray}\label{cq}
  \langle \Lambda_b(v',s')| j^{quark}_\mu(x=0)|\Lambda_b(v,s) \rangle &=&\int \frac{d^4 q}{(2 \pi)^4} \bar{\chi}(p') \gamma_\mu \chi (p) S_D^{-1}(p_2),\\
  \langle \Lambda_b(v',s')| j^{diquark}_\mu(x=0)|\Lambda_b(v,s) \rangle &=&\int \frac{d^4 q}{(2 \pi)^4} \bar{\chi}(p')\Gamma_\mu \chi (p) S_q^{-1}(p_1).
\end{eqnarray}
 Hence, we can calculate $g_1$ and $g_2$ as the following:
 \begin{eqnarray}
   g_1(\omega) &=& g_{1q}(\omega)-g_{1D}(\omega)\label{gg1} , \\
   g_2(\omega) &=& g_{2q}(\omega)-g_{2D}(\omega)\label{gg2} ,
 \end{eqnarray}
where $g_{iq}(\omega)$ and $g_{iD}(\omega)$ ($i=1,2$) are  from quark and diquark current contributions, respectively. 
The minus signs in Eqs. (\ref{gg1}, \ref{gg2}) are due to the relative charge between the quark and the diquark. So we have:

\begin{eqnarray}\label{qdq1}
\bar{u}(v',s')[g_{1q}(\omega) \gamma_\mu + g_{2q}(\omega) (v' + v)_\mu]u(v,s)&=&\int \frac{d^4 q}{(2 \pi)^4} \bar{\chi}(p') \gamma_\mu \chi (p) S_D^{-1}(p_2),
\end{eqnarray}
\begin{eqnarray}\label{qdq2}
\bar{u}(v',s')[g_{1D}(\omega) \gamma_\mu + g_{2D}(\omega) (v' + v)_\mu]u(v,s)&=&\int \frac{d^4 q}{(2 \pi)^4} \bar{\chi}(p') \Gamma_\mu \chi (p) S_q^{-1}(p_1).
\end{eqnarray}

\section{Numerical analysis}

\subsection{Solution of the BS wave functions}
In order to solve Equations (\ref{eig1}, \ref{eig2}), we define $M_{\Lambda_b}=m_b+m_D+E$ where $E$ is the binding energy. 
Taking $m_b=5.02$ GeV, $M_{\Lambda_b}=5.62$ GeV we have $ m_D+E=0.6$ GeV for $\Lambda_b$ \cite{Zhang-L}. 
We choose the diquark mass $m_D$ to be from $0.70$ to $0.80$ GeV for $\Lambda_b$. 
So the binding energy $E$ is from $-0.2$ to $-0.1$ GeV. 
The parameter $\kappa$ is taken to change from $0.02$ to $0.08$ GeV$^3$ \cite{Wu-HK}. 
Hence, for each $m_D$, we can get a best value of $\alpha_{seff}$ corresponding to a value of $\kappa$. 
Generally, $\tilde{f}_i~(i=1,2)$ should decrease to zero when $p_t \rightarrow + \infty$.
We change variables as the following:
\begin{eqnarray}
  p_t &=& \varepsilon + 3 \log \left[1+ 0.3 \frac{1+t }{1-t}\right],
\end{eqnarray}
where $\varepsilon$ is a small parameter in order to avoid divergence in numerical calculations, the range of $t$ is from $-1$ to $ 1$. Now we can use Gaussian quadrature method to solve Eqs. (\ref{eig1}, \ref{eig2}). 
Dividing the integration region into $n$ small pieces ($n$ is sufficiently large), the integral equations in Eqs. (\ref{eig1}, \ref{eig2}) become the following matrix equations:
\begin{eqnarray} \label{Meq1}
   f_{1 i} &=& A_{1 i j} f_{1 j}+ B_{1 i j} f_{2 j} +A_{2 i j} f_{1 j}+ B_{2 i j} f_{2 j},
\end{eqnarray}
\begin{eqnarray}\label{Meq2}
   f_{2 i} &=& A'_{1 i j} f_{1 j}+ B'_{1 i j} f_{2 j} +A'_{2 i j} f_{1 j}+ B'_{2 i j} f_{2 j}.
\end{eqnarray}

Comparing Eqs. (\ref{eig1}, \ref{eig2}) and (\ref{Meq1}, \ref{Meq2}), it is very easy to get the  matrices $A^{(')}_{(1,2)}$ and $B^{(')}_{(1,2)}$ (where $A$ and $B$ contain Jacobian determinates).
Solving matrix equations (\ref{Meq1}, \ref{Meq2}) we can get numerical solutions of the BS wave functions. In Table \ref{TB1}, we give the values of $ \alpha_{seff}$ for $m_D=0.70,~0.75,~0.80$ GeV for different $\kappa$.

\begin{figure}[!htb]
\begin{center}
\includegraphics[width=18cm]{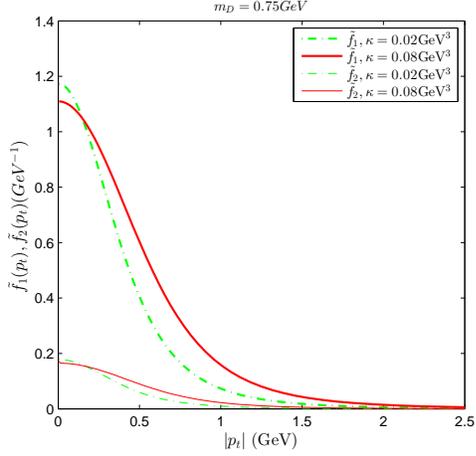}
\caption{(color online ) The BS wave functions for $\Lambda_b$ when $m_D=0.75$ GeV.}\label{FG1-bs}
\end{center}
\end{figure}

\begin{figure}[!htb]
\begin{center}
\includegraphics[width=18cm]{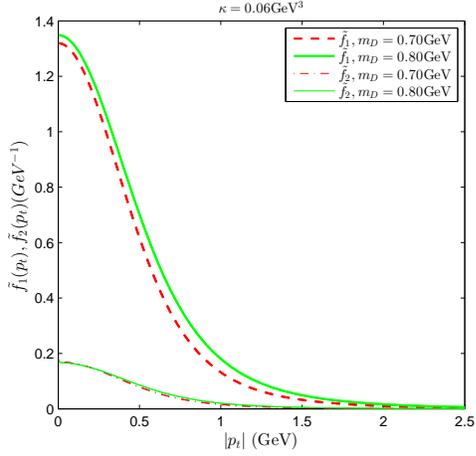}
\caption{(color online ) The BS wave functions for $\Lambda_b$ when $\kappa=0.06$ GeV$^3$.}\label{kp006}
\end{center}
\end{figure}

\begin{table}[!htb]
\centering  % 表居中
\begin{tabular}{l||c|c|c|c}  % {lccc} 表示各列元素对齐方式，left-l,right-r,center-c
\hline
  &$\alpha_{seff}(\kappa=0.02) $ &$\alpha_{seff}(\kappa=0.04 )$ &$\alpha_{seff}(\kappa=0.06 )$&$\alpha_{seff}(\kappa=0.08 )$ \\ \hline \hline% 在此行下面画一横线
$m_D=0.70$ &0.72 &0.76 &0.78 &0.80 \\         % \\ 表示重新开始一行
$m_D=0.75$ &0.76 &0.78 &0.80 &0.82 \\        % & 表示列的分隔线
$m_D=0.80$ &0.80 &0.82 &0.84 &0.86\\ \hline \hline
\end{tabular}
\caption{Values of $\alpha_{seff}$  for $\Lambda_b$ with different $m_D$ (GeV) and $\kappa~(\text{GeV}^3)$.}\label{TB1}
\end{table}
In Figs. \ref{FG1-bs} and \ref{kp006}, we plot $\tilde{f}_i~(i=1,2)$ depending on $|p_t|$.
 We can see from these figures that for different $\alpha_{seff}$ and $\kappa$, the shapes of BS wave functions are quite similar. 
All the wave functions decrease to zero when $|p_t|$ is larger than about $2.5$ GeV due to the confinement interaction.

\subsection{Calculation of EM form factors of $\Lambda_b$}

In order to solve Eq. (\ref{qdq1}), we use the following definitions:
\begin{eqnarray}
  \int \frac{d^4p}{(2\pi)^4}f_1^\prime(p^\prime)   f_1(p) S_D^{-1}(p_2)&=& k_0 \label{kk0},\\
  \int \frac{d^4p}{(2\pi)^4}f_1^\prime(p^\prime)  p_t^\mu f_2(p)  S_D^{-1}(p_2)&=& k_1 v^\mu +k_2 v^{ \prime\mu}\label{kk1},\\
 \int \frac{d^4p}{(2\pi)^4} f_2 ^ \prime(p^\prime) p_t^{\mu\prime} f_1(p) S_D^{-1}(p_2)&=& k_3 v^\mu +k_4 v^{\prime\mu}\label{kk5},\\
  \int\frac{d^4p}{(2\pi)^4} f_2^\prime(p^\prime) p^{\prime \mu}_t  p^{ \nu}_t  f_2(p) S_D^{-1}(p_2)& = & k_5 g^{\mu \nu }  +k_6 v^{\prime\mu} v^\nu + k_7 v^{\mu} v^{\prime\nu }\label{kk3},
\end{eqnarray}
where $k_i~(1=1,2,3...7)$ are functions of $\omega$. It is easy to prove 

\begin{eqnarray}
 k_1 &=& - \omega k_2,   \\
  k_4 &=& -\omega k_3 ,  \\
  k_6 &=&  0,  \\
  k_5 &=& - \omega k_7.
\end{eqnarray}
Then, we have:

\begin{eqnarray}
     k_0&=&\int \frac{d^4p}{(2\pi)^4}f_1^\prime(p^\prime)   f_1(p ) S_D^{-1}(p_2),  \label{k0} \\
  k_2 &=& \frac{1}{1-\omega^2} \int \frac{d^4p}{(2\pi)^4} f_1^\prime(p^\prime) {p_t \cdot v^\prime} f_2(p ) S_D^{-1}(p_2),  \label{k2} \\
  k_3 &=& \frac{1}{1-\omega^2} \int \frac{d^4p}{(2\pi)^4}  f_2^\prime(p^\prime) {p_t^\prime \cdot v} f_1(p ) S_D^{-1}(p_2),  \label{k3} \\
  k_5 &=& \frac{1}{3} \int \frac{d^4p}{(2\pi)^4}  f_2^\prime(p^\prime) {p_t^\prime \cdot p_t} f_2(p ) S_D^{-1}(p_2).\label{k5}
\end{eqnarray}

Define $\theta$ to be the angle between $p_t$ and $v^\prime_t$ where $v^\prime_t=v^\prime-(v\cdot v^\prime)v$, then we have 
\begin{eqnarray}
|v_t^\prime| &=& \sqrt{\omega^2-1} \label{vt},\\
 p_t \cdot v_t^\prime &=& - |p_t| |v_t^\prime | \cos\theta. \label{pvt}
\end{eqnarray}

Then we obtain the following relations:
\begin{eqnarray}
p_t \cdot v_t^\prime &=&- |p_t| \sqrt{\omega^2-1} \cos\theta \label{pvtt}, \\
p^\prime_t \cdot v &=& p_l(1-\omega^2) + |p_t|\omega \sqrt{\omega^2-1} \cos\theta +  m_D( \omega-1 )^2 \label{ptv}.\\
p_t \cdot p^\prime_t &=& (  p_l \omega - |p_t| \sqrt{\omega^2-1} \cos\theta -m_D \omega ) |p_t| \sqrt{\omega^2-1} \cos\theta  - |p_t| ^2  \label{pptt}.
\end{eqnarray}

Substituting Eqs. (\ref{SF}, \ref{SD}, \ref{vt} - \ref{pptt}) into  Eqs.(\ref{k0} - \ref{k5}), integrating $p_l$ and using the relation $\tilde{f}^{\prime}_{1(2)}=\int \frac{d p^{\prime}_l}{2 \pi}f^{\prime}_{1(2)}$, $k_i~(i=0,2,3,5)$ can be expressed in the terms of $\tilde{f}^{(\prime)}_{(1,2)}$.
Similarly, for solving Eq. (\ref{qdq2}), we repeat the above process with $S_F^{-1}$ being replaced by $S_D^{-1}(p_2)$, $k_{i}~(i=0,1,2...7)$ being replace by $k'_{i}$.
Furthermore. in Eqs. (\ref{ptv}, \ref{pptt}), we replace $m_D$ by $-m_b$. 
Finally, we obtain the following expressions for $g_{1q},~g_{2q},~g_{1D},~\text{and}~g_{2D}$:
\begin{eqnarray}
 g_{1q} &=&  k_0 -( \omega +1 )(k_2 + k_3) + \frac{k_5}{\omega}  \label{gq1}, \\
 g_{2q} &=& 2 \left(k_2 - \frac{k_5}{\omega}\right)  \label{gq2}, \\
 g_{1D} &=& 0 \label{gd1}, \\
 g_{2D} &=& k'_0 +2 (1-\omega) k'_2 +\left(2+\frac{1}{\omega}\right)k'_5  \label{gd2}.
\end{eqnarray}

Substituting Eqs. (\ref{g1q}, \ref{g2q}, \ref{gg1}, \ref{gg2}) into Eqs. (\ref{gem1}, \ref{gem2}) and considering the diquark contribution the EM form factors $G_E$ and $G_M$ can be written as 
\begin{eqnarray}
  G_E &=& g_{1q}- 2 \omega (g_{2q}-g_{2D}) \label{ge1} ,\\
  G_M &=& g_{1q}+ 6 (g_{2q}-g_{2D}) \label{gm1}.
\end{eqnarray}

\begin{figure}[!htb]
  \centering
  \includegraphics[width=6.5cm]{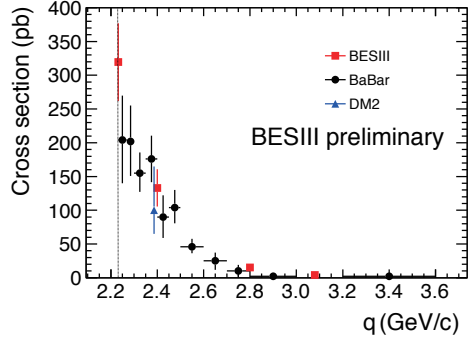}
  \caption{(color online) $\Lambda$ effective form factor (Data are taken from Ref. \cite{BESIII}).}\label{data}
\end{figure}

\begin{figure}[!htb]
  \centering
  \includegraphics[width=6.5cm]{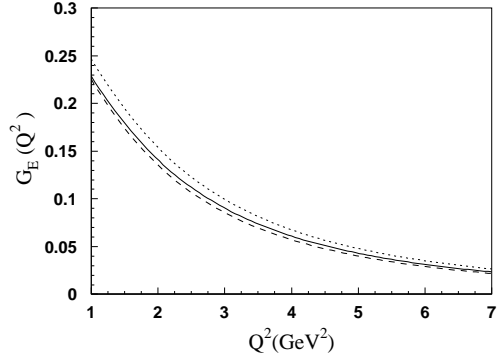}
  \caption{ $Q^2$-dependence of the electric form factor of  $\Lambda$ (This figure is taken from Ref. \cite{YL-L}).}\label{LYLE}
\end{figure}

According to the recent experimental data of BESIII \cite{BESIII} shown in Fig \ref{data}, the EM form factors of $\Lambda$ have a very large peak at small $Q^2$. 
In Ref. \cite{YL-L} the electric form factor of $\Lambda$ depends on $Q^2$ from $1 \sim 7$ GeV, corresponding to $\omega$ from $1.5$  to $4$, and the result is shown in Fig. \ref{LYLE}.

\begin{figure}[!htb]
\begin{center}
\includegraphics[width=18cm]{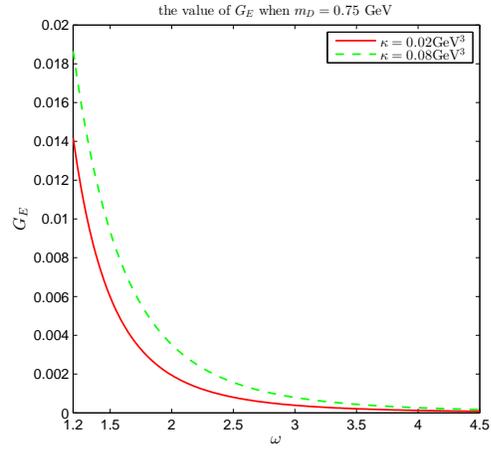}
\caption{(color online ) $\omega$-dependence of the electric form factor of $\Lambda_b$ for  $m_D=0.75$ GeV and different values of $\kappa$.}\label{GE75}
\end{center}
\end{figure}

\begin{figure}[!htb]
\begin{center}
\includegraphics[width=18cm]{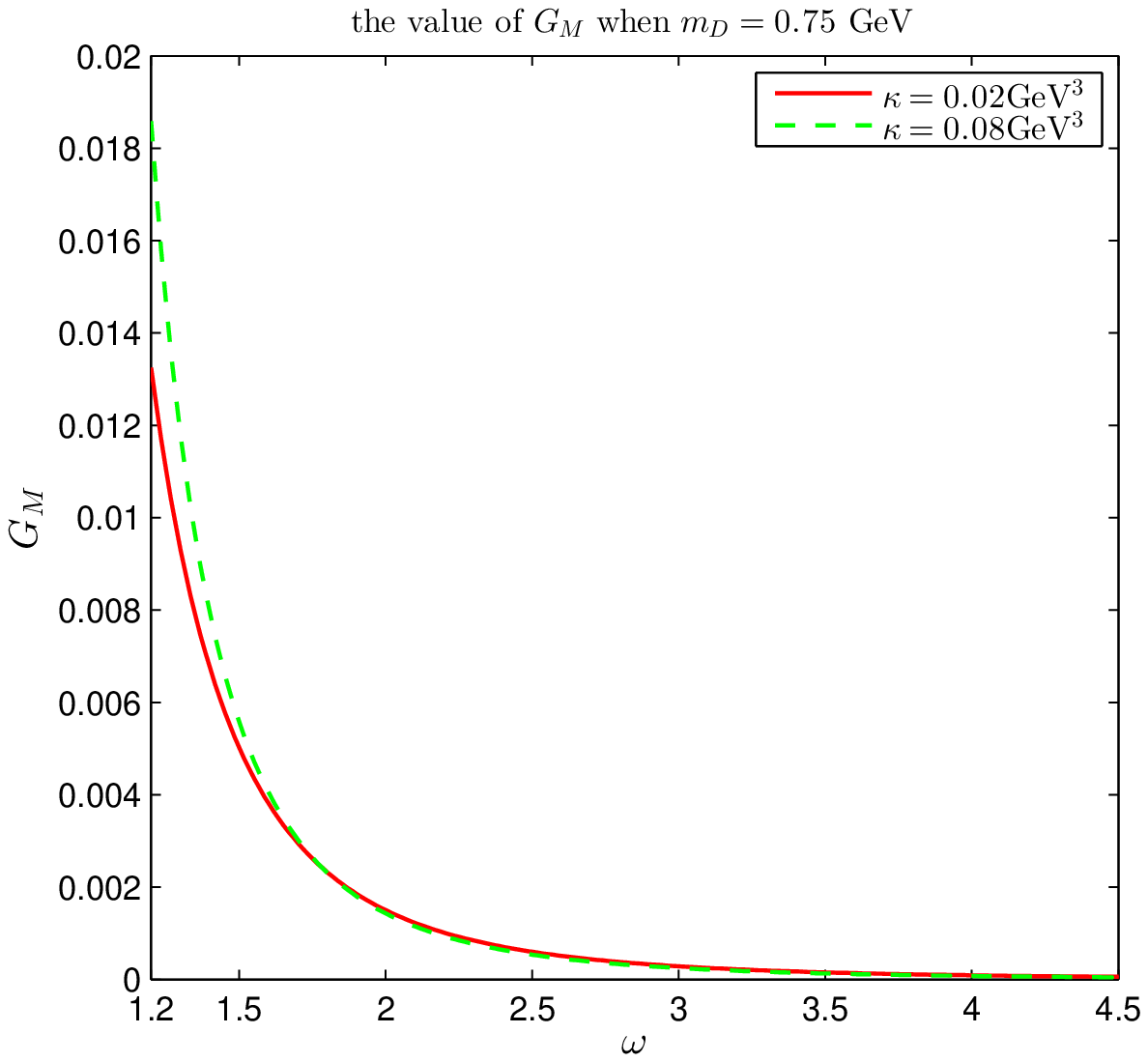}
\caption{(color online ) $\omega$-dependence of the magnetic form factor of $\Lambda_b$ for $m_D=0.75$ GeV and different values of $\kappa$.}\label{GM75}
\end{center}
\end{figure}

\begin{figure}[!htb]
\begin{center}
\includegraphics[width=18cm]{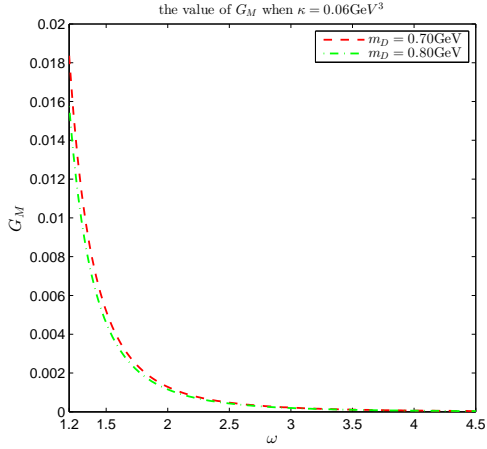}
\caption{(color online ) $\omega$-dependence of the magnetic form factor of $\Lambda_b$ for $\kappa=0.06$ GeV$^3$ and different values of $m_D$.  }\label{GMk6}
\end{center}
\end{figure}

\begin{figure}[!htb]
\begin{center}
\includegraphics[width=18cm]{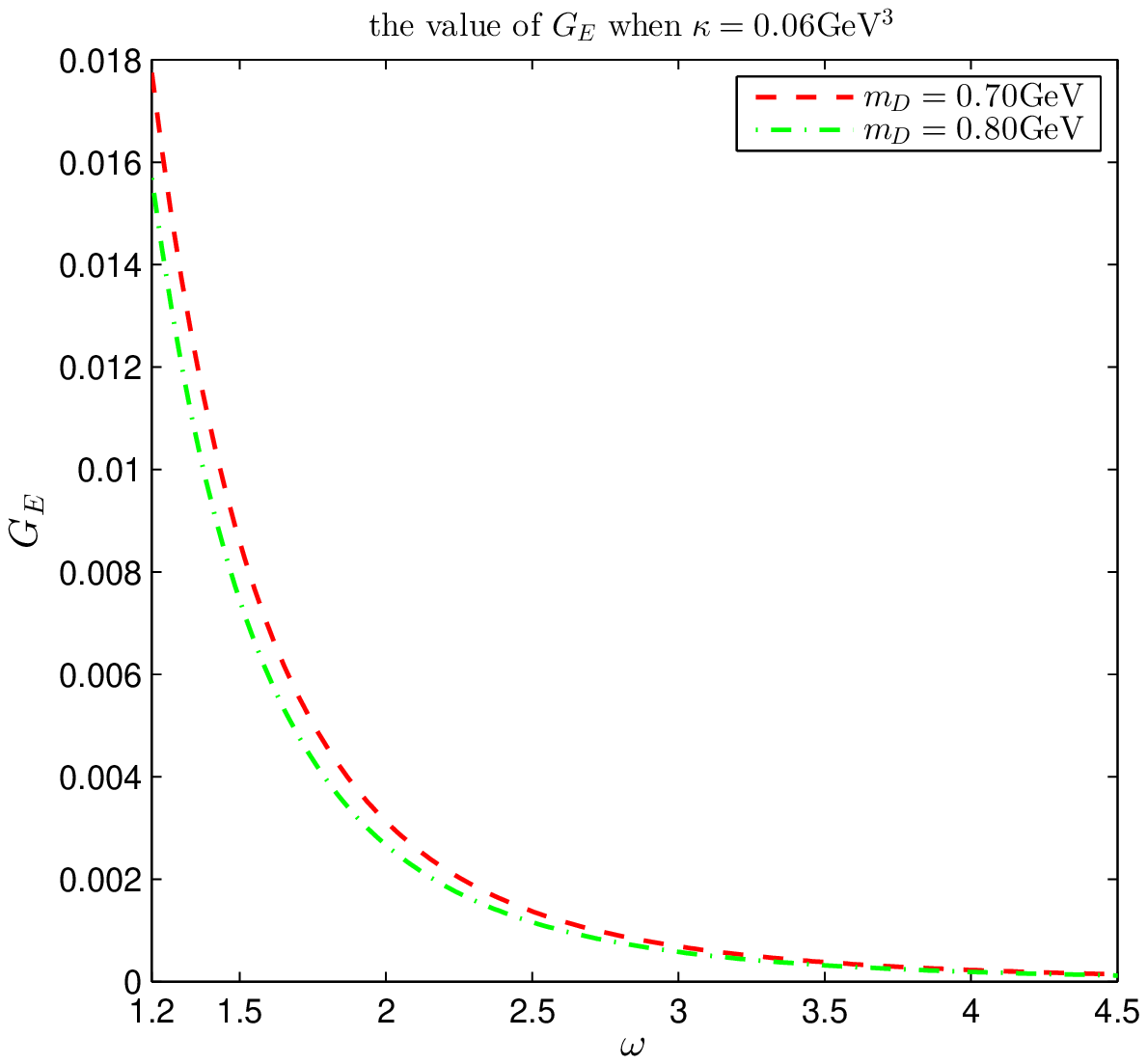}
\caption{(color online ) $\omega$-dependence of the electric form factor of $\Lambda_b$ for $\kappa=0.06$ GeV$^3$ and different values of $m_D$. }\label{GEk6}
\end{center}
\end{figure}

With the normalization condition Eq. (\ref{Nor}), solving Eqs. (\ref{qdq1}, \ref{qdq2}), we give the EM form factors $G_E(\omega)$ and $G_M(\omega)$ in Figs. \ref{GE75}-\ref{GEk6}.

From Figs. \ref{GE75}-\ref{gek6}, we find that for different $m_D$ and $\kappa$, the shapes of $G_E$ and $G_M$ are similar. 
In the range of $\omega$ from $1.5$ to $4$, this trend is similar to $\Lambda$, but changing more quickly than $\Lambda$.
From these figures, we also find that $G_M$ decreases more rapidly than $G_E$ as $\omega$ increases.

In the dipole model, $G_M(Q^2)=\frac{\mu}{(1+Q^2/m_0^2)^2}$, $\mu \propto 1/M$ (M is the mass of baryon) corresponds to the baryon  magnetic moment and $m_0=\sqrt{0.89}$GeV  is a parameter \cite{HMQ}.
There is no data for EM form factors of $\Lambda_b$ at present.
However, for different baryons (such as a and b) the ratio of $|G_E|$ and $|G_M|$, $RM$, should be of order $M_b/M_a$, i.e.

\begin{equation}\label{guji}
RM=|\frac{G_{Ma} }{ G_{Mb}}| \sim \frac{M_b}{M_a}.
\end{equation}

For $\Lambda$ and $\Lambda_b$, $RM$ is about $0.194$ in the dipole model. From Ref. \cite{YL-L} we know that the magnetic form factor of $\Lambda$ decreases faster than that in the dipole model. 
So, we expect the real value of $RM$ could be about $10^{-2} \sim 10^{-1}$. 
In the range of $\omega$ from $1.5$ to $4.5$ our result for $|G_{M\Lambda_b}|$ varies from about $0.007$ to $0$ and in Ref. \cite{YL-L} $|G_{M\Lambda}|$ varies from about $0.38$ to $0$. Their ratio agrees roughly with our expectation.

\begin{figure}[htb]
\begin{center}
\includegraphics[width=18cm]{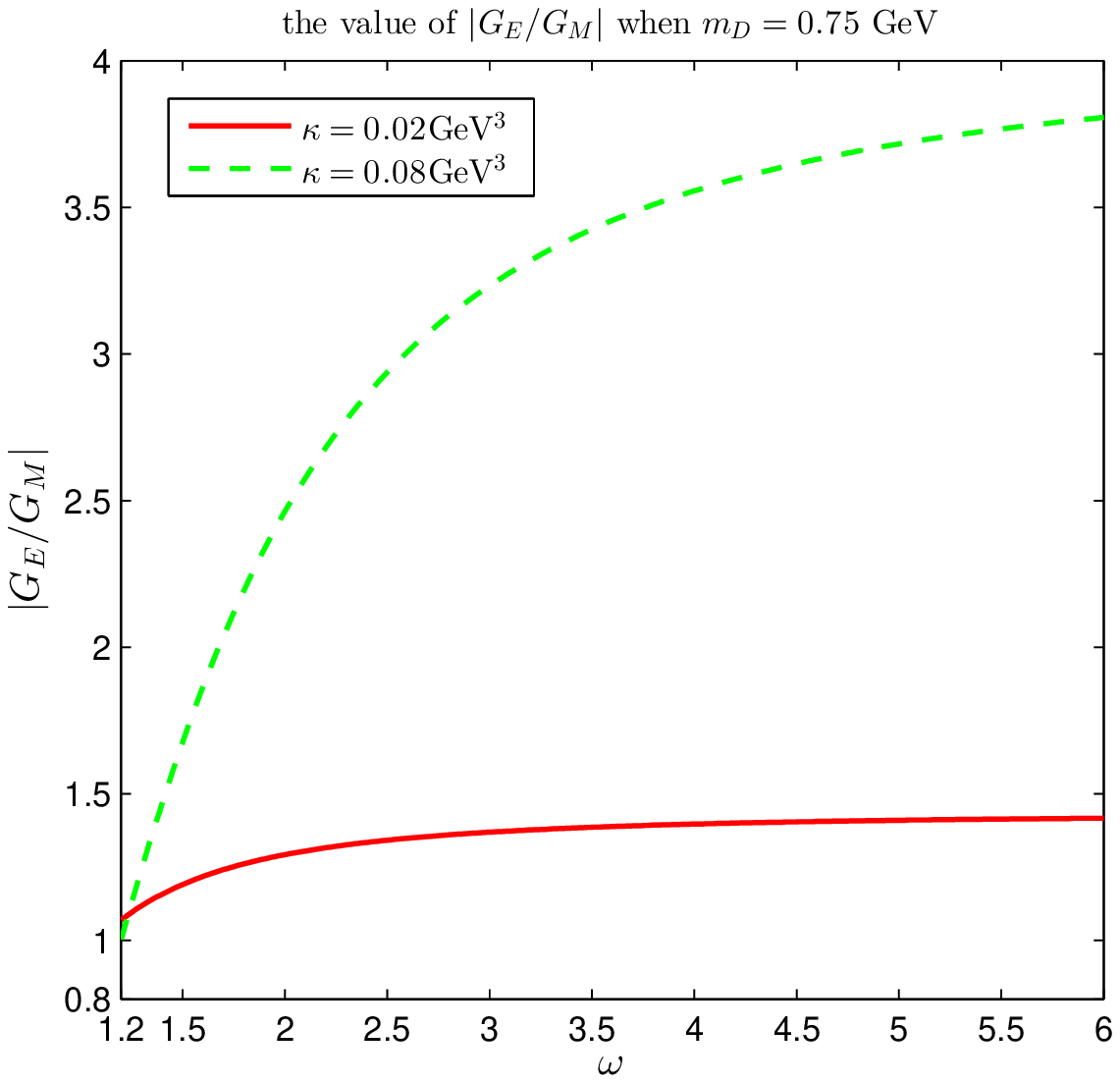}
\caption{(color online ) $\omega$-dependence of the form factor ratio $|\frac{G_E}{G_M}|$ for $\Lambda_b$. The results are for $m_D=0.75$ GeV and different values of $\kappa$. }\label{GEM75}
\end{center}
\end{figure}

\begin{figure}[!htb]
\begin{center}
\includegraphics[width=18cm]{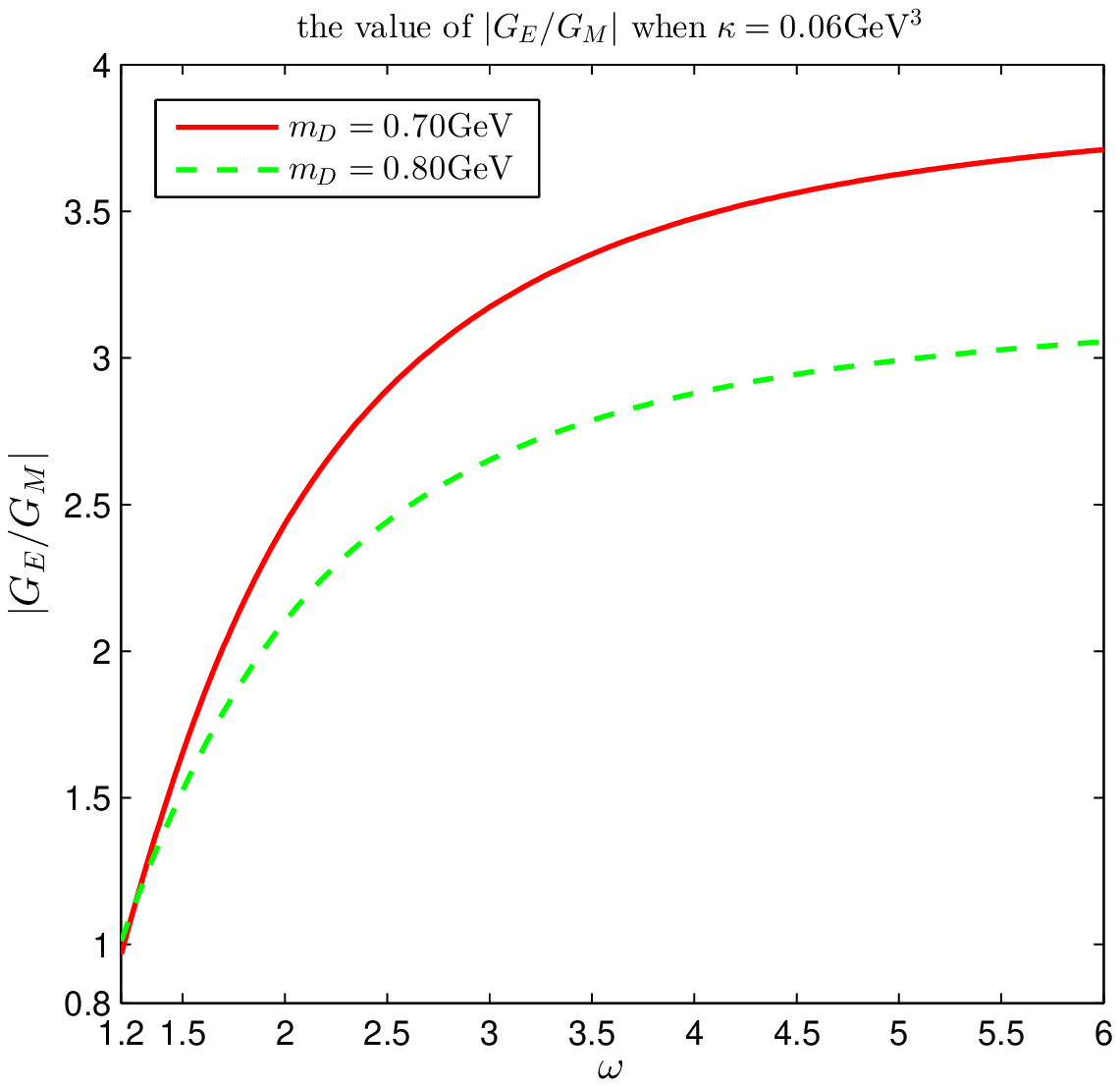}
\caption{(color online ) $\omega$-dependence the form factor ratio $|\frac{G_E}{G_M}|$ for $\Lambda_b$. The results are for $\kappa=0.06$ GeV$^3$ and different values of $m_D$. }\label{GEM80}
\end{center}
\end{figure}

The form factor ratio $|\frac{G_E}{G_M}|$ is often used to describe the angular distributions and model dependence of the detection efficiency \cite{B. Aubert}. 
It can be directly measured in experiments. 
The results for the form factor ratio of $\Lambda$ from BaBar Collaboration is \cite{B. Aubert},
\begin{eqnarray}
  |\frac{G_E}{G_M}| &=& 1.73^{+0.99}_{-0.57} ~\text{for} ~(2.23-2.4 \text{GeV}), \label{ratio1}\\
  |\frac{G_E}{G_M}| &=& 0.71^{+0.66}_{-0.71} ~\text{for} ~(2.40-2.80 \text{GeV}), \label{ratio2}
\end{eqnarray}
where the data (\ref{ratio1}) are fitted from $\Lambda \bar{\Lambda}$ threshold to $2.4$ GeV and the data (\ref{ratio2}) are fitted from $2.40$ GeV to $2.80$ GeV. 
Theoretically, at the threshold energy, the form factors ratio is $1$ \cite{B. Aubert}.
We give the form factor ratio $|\frac{G_E}{G_M}|$ for  $\Lambda_b$ in Figs. \ref{GEM75} and \ref{GEM80}. 

From Figs. \ref{GEM75} and \ref{GEM80}, we can see that for different $m_D$ and $\kappa$ the value of $|\frac{G_E}{G_M}|$ approaches $1$ when $\omega$ approaches $1$. 
It agrees with the theoretical result.
This means that Eq. (\ref{Nor}) is a good approximation and can be used to normalize the BS wave functions for the heavy baryon instead of Eq. (\ref{BSNOR}).

\section{summary and discussion}

Nowadays, more and more data about $\Lambda_b$ have been collected in experiments. 
In the quark-diquark picture, $\Lambda_b$ is regarded as a bound state of a heavy $b$-quark and a light scalar diquark based on the fact that the light degrees of freedom in $\Lambda_b$ have good spin and isospin quantum numbers. 
In this picture, we established the BS equation for $\Lambda_b$. 
Then we solved the BS equation numerically by applying the kernel which includes the scalar confinement and the one-gluon-exchange terms. 
Then, we calculated the EM form factors of $\Lambda_b$, and compared the results with those of $\Lambda$. 
It was found that the EM form factors of $\Lambda_b$ have a large peak at the threshold energy and the peak is much steeper than $\Lambda$. 
For different values of $m_D$ and $\kappa$ the EM form factors of $\Lambda_b$ change in the range   $0.018 \sim 0 $ as $\omega$ changes form $1.2$ to $4.5$.
The ratio of $|G_E|/|G_M|$ approaches 1 at small $\omega$.
This agrees with theoretical predictions.
We found that the normalization relation (\ref{Nor}) is a good approximation, which can replace the normalization relation of the BS wave function for the heavy baryon, Eq. (\ref{BSNOR}).

Depending on the parameters $m_D ~\text{and}~ \kappa$ in our model, our results vary in some ranges.
We studied the uncertainties for $G_E$ and $G_M$ that can be caused by $m_D$ and $\kappa$ and find that these uncertainties are at most about $48\%$ due to $\kappa$ and $18\%$ due to $m_D$. 
Our results need to be tested in future experimental measurements.
In the future, our model can be used to study other baryons such as the proton, the neutron, $\Lambda$ and $\Lambda_c$.

%%%%%%%%%%%%%%%%%%%%%%%%%%%%%%%%%%%%%%%%%%%%%%%%%%%%%%%%%
\acknowledgments
This work was supported by National Natural Science Foundation of China under contract numbers 11275025 and 11575023.
%%%%%%%%%%%%%%%%%%%%%%%%%%%%%%%%%%%%%%%%%%%%%%%%%%%%%%%%
%\appendix
%%%%%%%%%%%%%%%%%%%%%%%%%%%%%%%%%%%%%%%%%%%%%%%%%%%%%%%%%%%%%%%%%%%%%%%%%%%%%

%%%%%%%%%%%%%%%%%%%%%%%%%%%%%%%%%%%%%%%%%%%%%%%%%%%%


\begin{thebibliography}{99}

\bibitem{Arrington} J. Arrington, C. D. Roberts, and J. M. Zanotti, J.\ Phys.\ G {\bf 34}, 523 (2007).
\bibitem{Perdrisat} C. F. Perdrisat,  V. Punjabi, and M. Vanderhaeghen, Prog.\ Part.\ Nucl. Phys. \ {\bf 59}, 694 (2007).


\bibitem{Walk} R. C. Walker $et$ $al.$, Phys.\ Rev.\ D {\bf 49}, 5671 (1994); L. Andivahis $et$ $al.$, Phys.\ Rev.\ D {\bf 50}, 5491 (1994); M. E. Christy $et$ $al.$ (E94110 Collaboration), Phys.\ Rev.\ C {\bf 70}, 015206 (2004).
\bibitem{Arr} J. Arrington, Phys.\ Rev.\ C {\bf 68}, 034325 (2003).
\bibitem{Bost}I. A. Qattan $et$ $al.$, Phys.\ Rev.\ Lett {\bf 94}, 142301 (2005); P. Bourgeois $et$ $al.$, Phys.\ Rev.\ Lett {\bf 97}, 212001 (2006).
\bibitem{Lung} G. Kubon $et$ $al.$, Phys.\ Lett.\ B 524, {\bf 26} (2002).
\bibitem{VoL}J. Volmer $et$ $al.$ [The Jefferson Lab $\text{F}_{\Pi}$ Collaboration], Phys.\ Rev.\ Lett {\bf 86}, 1713 (2001).
\bibitem{Horn}T. Horn $et$ $al.$ [$\text{F}_{\Pi}$ Collaboration], Phys.\ Rev.\ Lett {\bf 97}, 192001 (2006)
\bibitem{Tade}V. Tadevosyan $et$ $al.$ [Jefferson Lab $\text{F}_{\Pi}$ Collaboration], Phys.\ Rev.\ C {\bf75}, 055205 (2007).
\bibitem{Cau}T. Van Cauteren $et$ $al.$, Eur.\ Phys.\ J.\ A {\bf 20}, 283 (2004); T. Van Cauteren $et$ $al.$, ArXiv: nucl-th/0407017.
\bibitem{Kub}B. Kubis, T. R. Hemmert and U. G. Meissner, Phys.\ Lett.\  B {\bf 456}, 240 (1999); B. Kubis and U. G. Meissner, Eur.\ Phys.\ J.\  C {\bf 18}, 747 (2001).
\bibitem{GDK} G. Ramalho, D. Jido, and K. Tsushima. Phys. Rev. D {\bf85}, 093014 (2012).
\bibitem{Bourgeois} P. Bourgeois $et$ $al.$, Phys.\ Rev.\ Lett. {\bf 97}, 212001 (2006).
\bibitem{Jlab} M. K. Jones $et$ $al.$, Phys.\ Rev.\ Lett.\ {\bf84}, 1398 (2000); O. Gajou $et$ $al.$, Phys.\ Rev.\ Lett \ {\bf 88}, 092301 (2002).
\bibitem{Len} V. M. Braun, A. Lenz, N. Mahnke and E. Stein, Phys.\ Rev.\ D {\bf 65}, 074011 (2002); V. M. Braun, A. Lenz, and M. Wittmann, Phys.\ Rev.\ D {\bf 73}, 094019 (2006); A. Lenz, M. Wittmann, and E. Stein, Phys.\ Lett.\ B {\bf 581}, 199 (2004).
\bibitem{Col} P. Colangelo, A. Khodjamirian, CERN-TH/2000-296, BARI-TH/2000-394.
\bibitem{J.R.Green} J. R. Green, J.W. Negele, and A. V. Pochinsky. Phys.\ Rev.\ D {\bf 90}, 074507 (2014).
\bibitem{Fra} J. Franklin, Phys.\ Rev.\ D {\bf 66}, 033010 (2002).
\bibitem{YL-L} Y.-L. Liu and M.-Q. Huang. Phys. Rev. D {\bf79}, 114031 (2009).
\bibitem{HMQ} Y.-L. Liu, M.-Q. Huang, D.-W. Wang, Eur. Phys. J. C {\bf 60}, 593 (2009).

\bibitem{BESIII} C. Morales [BESIII Collaboration], AIP Conf.\ Proc.\  {\bf 1735}, 050006 (2016).
\bibitem{J-U} J. Haidenbauer, U. G. Meibner. Phys. Lett. B {\bf761}, 456 (2016).%%-22-%%
\bibitem{JXU} J. Haidenbauer, X. W. Kang, U. G. Meibner, Nucl. Phys. A {\bf 929} 102 (2014).
\bibitem{H.Meyer} H. Meyer, Phys. Lett. B {\bf337}, 37 (1994).
\bibitem{JKV1} J. Haidenbauer, K. Holinde, V. Mull, J. Speth, Phys. Lett. B {\bf291} 223 (1992).
\bibitem{JKV2} J. Haidenbauer, K. Holinde, V. Mull, J. Speth, Phys. Rev. C {\bf46} 2158 (1992).
\bibitem{A. De Ruijula} A. De Ruijula, H. Georgi and S.L. Glashow, Phys. Rev. D {\bf12}, 147 (1975).
\bibitem{G. Karl} G. Karl, N. lsgur and D. W. L. Sprung, Phys. Rev. D {\bf23}, 163 (1981).
\bibitem{F. Close} F. Close, An introduction to Quarks and Partons (Academic Press, London, 1979) p. 302; H. Meyer and P. J. Mulders, Nucl. Phys. A {\bf 528} 589 (1991).
\bibitem{Guo-XH} X.-H. Guo and T. Muta, Phys. Rev. D {\bf 54}, 4629 (1996).
\bibitem{M.Ansel} M. Anselmino, P. Kroll, B. Pire, Z. Phys. C. {\bf36}, 89 (1987).
\bibitem{GS} G. P. Lepage and S. J. Brodsky, Phys. Rev. D {\bf22}, 2157 (1980); S. J. Brodsky, G. P. Lepage, T. Huang, and P. B. MacKenzie, in Particles and Fields 2, edited by A. Z. Capri and A. N. Kamal (Plenum, New York, 1983), p. 83.

\bibitem{Zhang-L} L. Zhang and X.-H. Guo, Phys. Rev. D {\bf87}, 076013 (2013).
\bibitem{Liu Y} Y. Liu, X.-H. Guo, and C. Wang,Phys. Rev. D {\bf91}, 016006 (2015).
\bibitem{Wu-HK} X.-H. Guo and H.-K. Wu, Phys. Lett. B {\bf654}, 97 (2007).
\bibitem{E.Eichten} E. Eichten, K. Gottfried, T. Kinoshita, K. DLane, and T. M. Yan, Phys. Rev. D {\bf17}, 3090 (1978).
\bibitem{Weng-Mh} M.-H. Weng, X.-H. Guo, and A.W. Thomas, Phys. Rev. D {\bf83}, 056006 (2011).
\bibitem{Wei-Kw} X.-H. Guo and X.-H. Wu, Phys. Rev. D {\bf76}, 056004 (2007).
\bibitem{G.P} G. Peter Lepage and Stanley J. Brodsky. Phys. Lett. B {\bf 87}, 359 (1979).
\bibitem{GPL} G. P. Lepage and S. J. Brodsky, Phys. Rev. Lett. {\bf43}, 545 (1979); 43, 1625(E) (1979); Phys. Rev. D {\bf22}, 2157 (1980).
\bibitem{A.Efr} A. Efremov and A. Radyushkin, Phys. Lett. B {\bf94}, 245 (1980).
\bibitem{V.L.C} V. L. Chernyak and A. R. Zhitnitsky, JETP Lett. {\bf25}, 510 (1977); Phys. Rep. {\bf112}, 173 (1984).
\bibitem{I.G.A} I. G. Aznaurian, S. V. Esaibegian, K. Z. Atsagortsian, and N. L. Ter-Isaakian, Phys. Lett. B {\bf90}, 151 (1980); B {\bf92},371(E) (1980).

\bibitem{V.A.A} V. A. Avdeenko, S. E. Korenblit, and V. L. Chernyak, Sov. J. Nucl. Phys. {\bf33}, 252 (1981).

\bibitem{C.E.C} C. E. Carlson and F. Gross, Phys. Rev. D {\bf36}, 2060 (1987); N. G. Stefanis, Eur. Phys. J. C {\bf1}, 7 (1999); A. Duncan and A. H. Mueller, Phys. Lett. B  {\bf90}, 159 (1979); A. H. Mueller, Phys. Rep. {\bf73}, 689 (1981).
\bibitem{VCM} V. Punjabi1, C. F. Perdrisat, M. K. Jones, E. J. Brash, and C.E. Carlson. Eur. Phys. J. A {\bf51}, 1 (2015).
\bibitem{ST} S. Drell and T. M. Yan, Phys. Rev. L {\bf24}, 181 (1970).
\bibitem{V.Keiner} V. Keiner, Z. Phys. A {\bf354}, 87 (1996).
\bibitem{G-K} X.-H. Guo,  P. Kroll, Z. Phys. C {\bf 59}, 567 (1993).
\bibitem{M-V} M. Neubert, V. Rieckert, Nucl. Phys. B {\bf 382} 97 (1992).
\bibitem{M-B} M. Wirbel, B. Stech, M. Bauer, Z. Phys. C {\bf29} 637 (1985).
\bibitem{H-M} H. Leutwyler, M. Roos, Z. Phys. C {\bf25}, 91 (1984).
\bibitem{BJM} B. K$\ddot{o}$nig, J. G. K$\ddot{o}$rner, M. Kr$\ddot{a}$imer, P. Kroll, Phys. Rev. D  {\bf56}, 4282 (1997).%%%%%---45----%%%%%%
\bibitem{C.R.M} C. R. M$\ddot{u}$nz, J. Resag, B. C. Metsch,, H. R. Petry, , Phys. Rev. C {\bf52}, 2110 (1995).
\bibitem{B. Aubert} B. Aubert, BaBar Collaboration, $et$ $al.$, Phys. Rev. D {\bf76}, 092006 (2007).



\end{thebibliography}
\end{document}